\documentclass[times]{speauth}
\usepackage{graphicx}
\usepackage{caption}
\usepackage{subcaption}
\usepackage{textcomp}
\usepackage{enumitem}
\usepackage{color,soul}
\usepackage[colorinlistoftodos]{todonotes}
\setitemize{noitemsep,topsep=0pt,parsep=0pt,partopsep=0pt}
\usepackage{hyperref}

\def\volumeyear{2016}

\begin{document}

\runningheads{Sreekrishnan et al.}{Architectural Partitioning and
  Deployment Modeling on Hybrid Clouds}

\title{Architectural Partitioning and Deployment Modeling on Hybrid Clouds}
\author{Sreekrishnan Venkateswaran\affil{1} and Santonu
  Sarkar\affil{2}\corrauth}
\address{
\affilnum{1}IBM Corporation, India 
\affilnum{2}BITS Pilani K.K.Birla Goa Campus, Goa India
}

\corraddr{BITS Pilani K.K.Birla Goa Campus, Goa India-403726. Email:
  santonus@goa.bits-pilani.ac.in}

\begin{abstract}

The hybrid cloud idea is increasingly gaining momentum because it
brings distinct advantages as a hosting platform for complex software
systems. However, there are several challenges that need to be
surmounted before hybrid hosting can become pervasive and
penetrative. One main problem is to architecturally partition
workloads across permutations of feasible cloud and non-cloud
deployment choices to yield the best-fit hosting combination. Another
is to predict the effort estimate to deliver such an advantageous hybrid
deployment.

In this paper, we describe a heuristic solution to address the said
obstacles and converge on the ideal hybrid cloud deployment
architecture, based on properties and characteristics of workloads
that are sought to be hosted. We next propose a model to
represent such a hybrid cloud deployment, and demonstrate a
method to estimate the effort required to implement and sustain that
deployment. We also validate our model through dozens of case
studies spanning several industry verticals and record results
pertaining to how the industrial grouping of a software system can
impact the aforementioned hybrid deployment model.

\end{abstract}


\maketitle
\vspace{-6pt}
\section{Introduction}
\vspace{-2pt}
Cloud Computing~\cite{nist_cloud}, specifically Infrastructure as a
service (IaaS), is a paradigm that supplies on-demand compute
capability with consumptive billing~\cite{Assuncao2009}. The on-demand
capability consists of a set of infrastructure units over which software
systems can be deployed; it can be, for example, a data center
comprising of several virtual servers, physical servers, firewalls,
load balancers, and storage drives exclusive dedicated for the
software system. The deployment choice can also be a public cloud
service such as Amazon EC2~\cite{EC2} that sells on-demand compute
capability on an open market. However, a public cloud customer has
little control over the hardware and software infrastructure of the
compute resource that is provided. A private cloud
(e.g. Nebula~\cite{Fontan2008,Nebula}) on the other hand, is typically
smaller in scale and confined within an organization. Compared to a
public cloud, private cloud users have more control over the cloud
infrastructure.

Hybrid Cloud~\cite{Sotomayor2009} combines a set of public clouds,
private clouds, as well as high-performing bare-metal
infrastructure. It provides the benefit of compute on demand like a
public cloud; it also provides better overall control over the compute
infrastructure and offers superior performance like a private
cloud. Consider a large global enterprise with hundreds of
applications developed over several years hosted on geographically
scattered infrastructure. Such a deployment can become deficient due
to adhoc mix-and-match of hardware and software resources over a period of
time. If these applications can be transformed and redeployed on a 
hybrid cloud, the enterprise can realize efficiency and cost
savings that arise out of best fitment of workloads with the
underpinning infrastructure. This has been observed and highlighted in
a recent Gartner survey~\cite{Gartner2015}, which predicted that the
adoption of hybrid cloud as the preferred hosting model among
enterprises will triple from 2015 to 2017. However, the complexities
involved in adopting hybrid cloud into mainstream businesses are
manyfold. One important issue is to achieve a judicious
allocation of workloads across different infrastructure platforms, since inefficient partitioning of workloads across the heterogeneous systems
constituting the hybrid cloud can result in violation of the system's 
Service Level Agreements (SLAs). According to a Technology Business Research (TBR) market survey report~\cite{TBR2014}, there is a 32\% gap between expected and actual evolution of cloud-hosted workloads to hybrid environments across enterprises. This means that while the public and private cloud markets have come of age, hybrid cloud maturity has not yet kept expected pace due to various reasons.

In this paper, we propose a heuristic-driven methodology, developed
from several real-life cloud based deployments for customers across
industries, to achieve an advantageous partitioning of workloads across a
hybrid cloud. We further propose a model to represent this 
partitioned deployment and demonstrate how the complexity of such a
deployment can be estimated in terms of the required implementation
effort. 

This paper is organized as follows. Section~\ref{sec:problem} elaborates 
the problem description. In Section~\ref{sec:partition}, we first describe 
a construct called the Cloud Line of Isolation and Control (CLIC) that we 
use as a basic architecture partitioning mechanism. We then formulate a model
in Section~\ref{sec:workload} and Section~\ref{sec:predicting-effort} 
that measures hybrid complexity and predicts deployment effort of classes 
of workloads that are sought to be deployed. Section~\ref{sec:empiricaldata} 
presents results on how industrial characteristics affect hybrid complexity. 
Section~\ref{sec:casestudy} validates our proposed hybrid deployment 
model using case studies from four industries where we applied this
technique to determine the contours of the hybrid deployment. Section~\ref{sec:relwork}
describes the related work in this area. Finally we conclude the paper
and highlight future research directions.
\section{Problem Description}
\label{sec:problem}

While hybrid cloud offers benefits of deployment cost reduction,
superior performance, and improved security, a confluence of
heterogeneous compute platforms is seldom perfect; there are
several impediments to surmount before opportunities can be
exploited.

One major problem confronting evolution to hybrid environments is that 
complex business workloads that need to be deployed on hybrid clouds 
must be carefully partitioned. The partitioning technique must
identify the appropriate platform on which each workload must be
deployed; else the deployment will not correspond to minimalist complexity 
that will satisfy requirements, and hence will be inefficient and expensive. 
Hence the analysis and placement decision of workloads must be taken apriori. 
This paper is devoted to addressing this problem space. We will elaborate on 
this problem next.
\vspace{-6pt}
\subsection{Partitioning of Software Systems for Hybrid Environments: Major Challenges}
As alluded to earlier, hybrid environments are composed of
multiple cloud and traditional data centers; the cloud is typically
multi-vendor and heterogeneous, comprising of private, public and
community ("shared private") deployments. While the deployment of future IT systems will turn increasingly hybrid~\cite{Gartner2015}, the challenge is to arrive at a topology that results in the least cost of ownership. There
are many permutations of hybrid deployment architectures available
that offer varying degrees of control, isolation, security and
performance. One or more of these permutations is likely to offer the
most advantageous fitment depending on the characteristics of various
categories of component workloads.

The problem thus, is to design a deployment strategy for the service provider that
apportions different workloads of a business system across a hybrid
infrastructure that satisfies various non-functional requirements
related to performance, security and availability, at the minimum
cost of ownership. A cloud deployment, at a high level, comprises of the cloud stack (which we refer to as the ``managing environment''), and the application workloads that run on target virtual machines (which we call the ``managed environment''). In this paper we focus on the complexity and effort in constructing the managing environment.

An additional problem is to model the effort required to implement such a hybrid cloud based deployment, which 
can then be used as a prediction tool for techno-commercial decisions 
spanning architecture and business.

\section{Building a Hybrid Deployment Model: A Structured Approach to Hybrid Partitioning} 
\label{sec:partition}
Software systems have different requirements on isolation and
control. Some workload components need heavy {\em isolation} and require high {\em architectural control} compared to others.

The desire for isolation translates to tolerance for other co-located
tenants. The degree of required isolation is a function of regulatory
compliance requirements, performance guarantees, and security. For
instance, it is a common perception that security is more compromised
if sensitive workloads are hosted off-premise, which translates to a 
preference for storing sensitive data close to the client’s office, 
i.e. a requirement for stronger isolation

Architectural control, on the other hand, reflects the ability to
customize the construction of the cloud management system. It may be
noted that the term architectural control in the context of
traditional software architecture~\cite{Aldrich2014} implies a set of constraints
that an architect imposes on a design so as to ensure that the
architectural decisions are met. In this context, the degree of
required architectural control depends on the quantum of bespoke
modifications that need to be incorporated into the cloud
infrastructure. It could arise out of the need to exercise code level
control on the cloud managing stack, the necessity to handle frequent
change requests, or the need to perform heavy duty integration with complex tools like sophisticated authentication systems, service monitoring and management tools.

Note that the stipulation for a heavy isolation does not imply the
necessity of high architectural control, and vice versa. Therefore, we
can treat them as orthogonal entities.

\subsection{Cloud Line of Isolation \& Control}
In this section, we define the Cloud Line of Isolation and Control
(CLIC) of a software system (comprising of a set of workloads) as the boundary that separates those
workload components that demand strong isolation and control, from
those that have relatively relaxed requirements.
\begin{figure} 
\centering 
\includegraphics[scale=0.47]{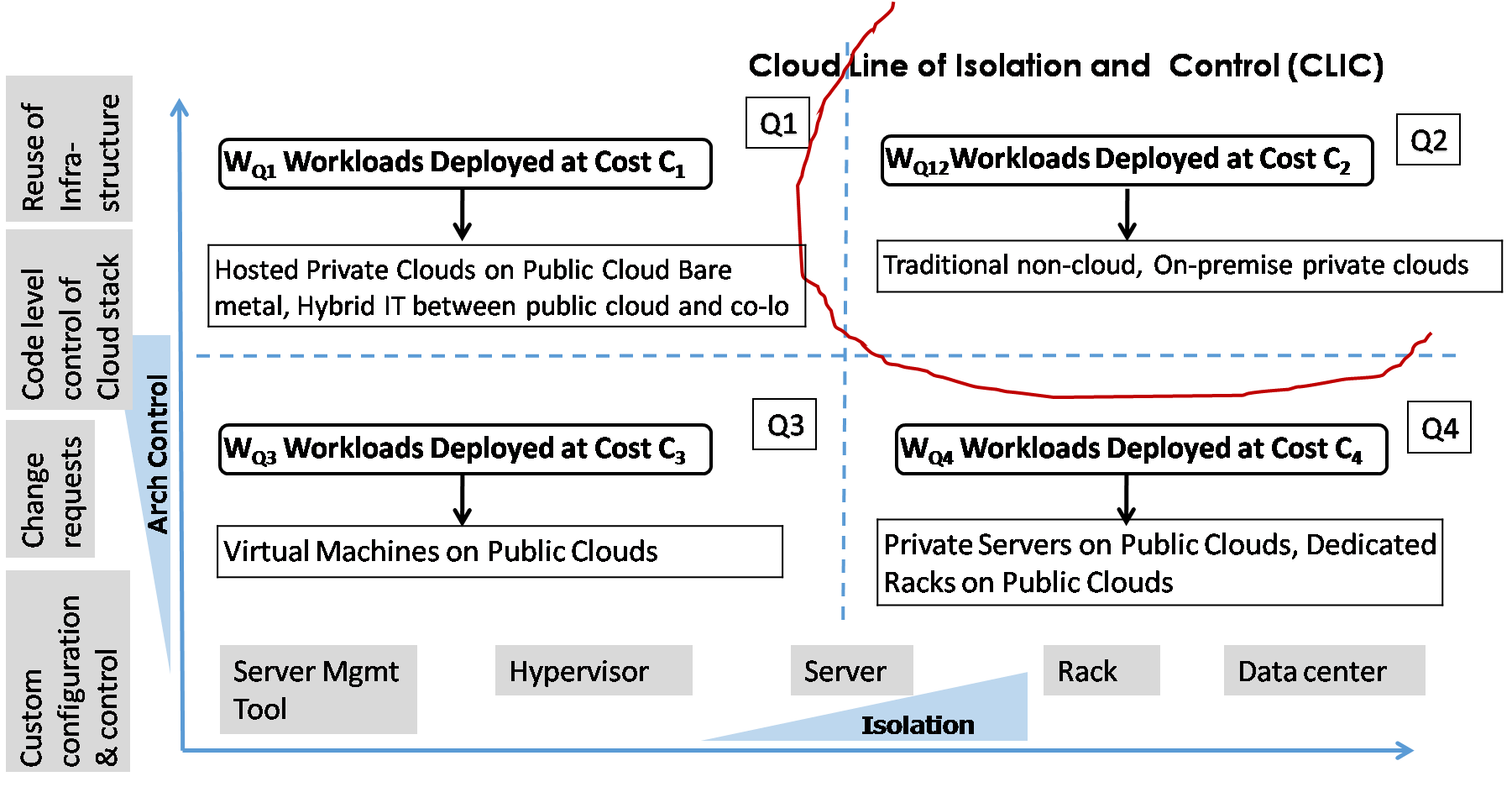}
\caption{Hybrid Cloud Components Mapped to Zones of Varying Isolation \& Control}
\label{fig:clic}
\vspace{-10pt}
\end{figure}
Before explaining the CLIC and the motivation behind introducing it,
we need to explain the cloud deployment graph depicted in
Figure~\ref{fig:clic}. This graph plots the following variables:\\
\indent
The X-axis traces the degree of isolation required by the software
systems that are to be deployed. Assume that a set of infrastructure
resource elements $R_{i,(0<i<N)}$ are hosting virtual machines for $M$
clients $C_{j,(0<j<M)}$. Let each infrastructure resource element
$R_i$ host $K$ virtual machines $V_1$ to $V_k$. Let these $K$ virtual
machines run a set of workloads. If all these $K$ virtual machines are
owned by the same client $C_j$, the workloads hosted on these machines
are said to be isolated. If $R_i$ is a physical server, the isolation is
at the host level; if $R_i$ is a rack, it is a stringent rack level
isolation; if $R_i$ implies a data center, then the isolation is even
harder because the hosting facility itself needs to be dedicated for
the client in question. However, if no such tenancy requirements
exist, there is a little isolation demand on the hosted workload. Note
that the question of isolation triggers the larger “shared versus
dedicated” question and not the “on-premise versus off-premise”
question.

The Y-axis grades the degree of architectural control, which
is the quantum of customization needed to build the cloud. The need
for architectural control can range from merely changing certain
configuration parameters of a standard cloud like AWS, to introducing
deep code-level changes inside the cloud stack.

The set of feasible deployment models pertaining to each region is the
third aspect, examples of which are covered inside the boxes attached to each
quadrant.

The workloads of a client are first plotted on this cloud deployment
graph based on their requirements on isolation and architectural
control. Next, the CLIC line is used to help segment the graph into
quadrants, thus separating those workloads with high isolation and
architectural control requirements from the rest. The CLIC helps to
quickly visualize a feasible high level approach to designing the
workload deployment architecture. The workload components to the left
of the CLIC are further trisected into three subcategories based on
their isolation and control requirements.

\subsubsection{Deployment Options}
As can be seen from Figure~\ref{fig:clic}, there are various feasible
deployment options on both sides of the CLIC (see the text boxes
embedded in the four quadrants) that are an assemblage of clouds and
traditional non-cloud based infrastructure. To explain this, we take a
quick tour of the graph.

We start with Quadrant \#3, the most relaxed in terms of desire for
architectural control, and demand for isolation. Public virtual
instances on multiple public clouds are generally sufficient to host
these workload components. VMs that host these portions of the overall
workload are provisioned on physical servers that are shared with
multiple customers. The cloud provider manages the hypervisor; hence
clients {\em do not have} control over the environment other than on
their virtual instances.

Quadrant \#4 is harder in terms of isolation needs, but 
architectural control continues to be relinquished to a large
extent. An example hosting route is via private virtual machines on a
public cloud, where there is isolation at the server level. A private
VM is deployed on a single-tenant physical server. No other customer
instance gets provisioned on that same physical server, thereby
assuring that this VM does not share resources with other
clients. The virtualization is controlled by the provider; the customer 
whose instances are provisioned on the dedicated server will not be able to 
exert control on the managing environment.

Next is Quadrant \#1, the region where workloads have low isolation
requirements, but the client is not ready to relinquish architectural
control. A deployment architecture that satisfies this is {\em hosted 
private clouds}. Such private clouds are owned, managed, and
operated by external providers. An example of a hosted private cloud
provider is IBM Bluebox~\cite{blueboxcloud}. Several single-tenant hosted 
options exist: off-premise private clouds, private clouds
deployed on-premise, or private clouds deployed in a collocation space
attached to the data center that houses a public cloud where other
workload components of the system reside.

Finally, we arrive at Quadrant \#2, the most stringent of the four
zones. It falls on the right side of the CLIC, and this usually calls
for on-premise private clouds or traditional IT hosting. However,
depending on the degree of required isolation, one option is to
use physical servers on demand in a public cloud. These are
single-tenant bare metal servers on multi-tenant public clouds
completely dedicated to a customer; no part of the server resources
will be shared with other customers. Hypervisors do not come attached
with the physical servers, but the client has the option to virtualize
purchased servers and also to stand up a private cloud on top of the
bare metal servers, which provides a high degree of
architectural control over the deployment.

\subsection{Modeling the Hybrid Complexity Of Workloads}
\label{sec:workload}

We now introduce the {\em degree of hybrid complexity of the best-fit
deployment} of a software system based on workload
characteristics. Encapsulating the complexity of the ideal hybrid
deployment of a software system in the form of a single metric is
important due to two reasons: it becomes possible to represent a
deployment\textquotesingle s complexity quantitatively, which can
constitute an input for architectural and business decisions; further
it allows comparing the relative hybrid deployment intricacies of two
or more software systems and hence helps in estimating the
implementation effort and steady state delivery effort post service
commencement. We will use the symbol $H(w)$ to refer to the aforementioned degree 
of hybrid complexity.

\subsubsection{Empirical Study}
\label{sec:empiricaldata}
\begin{table} [ht]
\centering
\caption{Empirical Observations of Effort Estimates to Deploy Singular Clouds} \centering
\tabsize
\begin{tabular}{|c|p{1.4in}|p{0.5in}|p{0.5in}|p{0.6in}|p{1.4in}|} 
\toprule
\# & Client's IT Environment & Quadrant (in our proposed model) & \# of Client Deals Analyzed & Average $C_i$ (Person Months) & Remarks \\
\midrule
1 & All workloads deployed on hosted private clouds (OpenStack atop IBM SoftLayer Cloud bare metal or VMware vRealize Cloud atop IBM SoftLayer Cloud bare metal) & 1 & 14 & 50 & Much more complex than \#3 or \#4, but savings accrued from cloud driven automation developed by leveraging available architectural control\\ 
2 & All workloads deployed on-premise on non-Cloud environments (core banking on main frames or industrial backend software on high performance physical servers) & 2 & 15 & 120 & The environment not being software defined, could not exploit cloud-like efficiencies, so labor to run the data center was high. 'Toughest' of the quadrants.\\ 
3 & All workloads deployed on public clouds (Amazon AWS, Microsoft Azure or IBM  SoftLayer) & 3 & 20 & 22 & Retail cloud hosting with no architectural control and minimum isolation. Most relaxed in terms of requirements\\ 
4 & All workloads deployed on private VMs on public clouds (Amazon AWS dedicated instances or IBM SoftLayer private instances) & 4 & 11 & 45 & Infrastructure cost is high even though the hosting is similar to \#3\\ 

\bottomrule
\end{tabular}
\label{tab:data4model}
%\vspace{-5pt}
\end{table}
In order to characterize hybrid complexity, we observed 60 client cloud deployments across different industries where the IT infrastructure was hosted on environments where the architecture was singular and not hybrid. We then categorized these deals into 4 sets based on the quadrants of Figure~\ref{fig:clic} in which they resided. For each of the four categories, we calculated the average cost ($C_i$) to build and run the cloud managing environment. The deployment cost $C_i$ comprises of:
\begin{enumerate}[noitemsep,topsep=0pt,parsep=0pt,partopsep=0pt]
\item the cost of the BoM (Bill of Material) or the 'raw materials' to deploy the solution; in this case the hardware infrastructure and software/middleware licensing costs to implement the cloud stack 
\item the cost of the effort required to implement the overall hybrid cloud stack
\item the cost of managing the cloud stack during steady state operations for a one-year period, and
\item the cost of automating the process of making the target environment manageable. For example, suppose that a particular business system for a given industry sector, demands improved dependability, through the infrastructure monitoring tool Nagios\footnote{https://www.nagios.org} post provisioning. In such a case, the cloud stack needs to ensure that every VM has the Nagios agent running as a part of the deployment and sustenance exercise. Similarly, the service provider may also needs to take care of patching, ticketing, anti-virus etc. Also, if provisioning needs to trigger an approval workflow, the cloud stack needs to have that implemented.
\end{enumerate}
All costs (including infrastructure charges) are depicted in units of person months of labor for ease of comparative calculations. This cost will enjoy a return on investment in the form of savings that will accrue in managing the target infrastructure due to introduction of cloud efficiency and associated automation. 
 
Note that we are considering the cloud stack, which is the managing environment and not the managed environment, while computing the cost. So, in the case of public clouds (row \#3 and row \#4 of Table~\ref{tab:data4model}), the semantics of associated effort is slightly different from that for private clouds. For the former, the effort is a combination of the labor required to provision the target infrastructure via the public cloud provider\textquotesingle s portal, develop any custom integration, and resolve cloud specific steady state issues. Thus, row \#3 of Table~\ref{tab:data4model} does not include any BoM cost in column \#5 since the managing environment is built by the provider; row \#4, however, includes a BoM component that is equivalent to the differential between dedicated and shared virtual instances on the deployed public cloud.

The result of our observations is a measure of how much each Quadrant of Figure~\ref{fig:clic} contributes to the overall complexity of a hybrid deployment. The cost depicted in column 5 of Table~\ref{tab:data4model} reveals a 2:5:1:2 ratio across the four quadrants $Q_1$:$Q_2$:$Q_3$:$Q_4$. 
 
This ratio essentially implies that the deployment cost of a workload is the least in Quadrant \#3; costs nearly double when deployed to Quadrant \#4; and costs five times more when deployed to Quadrant \#2.\\

\begin{figure} 
\centering 
\includegraphics[scale=0.8]{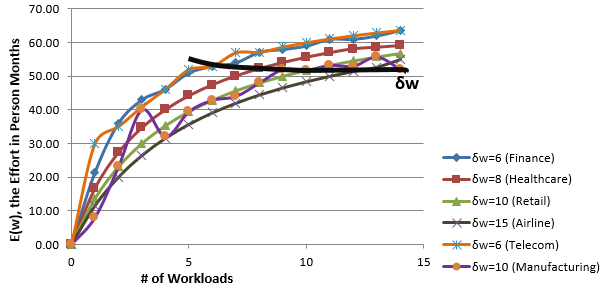}

\caption{$\delta_w$ across industries (essentially $W_i$ beyond which effort starts plateauing)}
\label{fig:delta-graph}
\vspace{-5pt}
\end{figure}

We also performed another set of studies to observe the movement of deployment cost ($C_i$) as a function of the number of workloads in a given Quadrant. For this purpose, we collected 90 cloud deployment solutions as follows:
\begin{itemize}[noitemsep,topsep=0pt,parsep=0pt,partopsep=0pt]
\item The chosen client cloud solutions were equally distributed across 6 industries, thus $15$ deployments each from finance, healthcare, retail, airline, telecom, and manufacturing.
\item They were chosen such that for solution $k$ ($k~=~1~to~15$) in each of the 6 industries, the number of workloads to be deployed in Quadrant \#1 was $k$.

\item  We then recorded the observed person month effort (we call it $E(w)$) to deploy the Quadrant \#1 cloud and sustain it for one year.

We normalized this cost as follows:
\begin{itemize}[noitemsep,topsep=0pt,parsep=0pt,partopsep=0pt]
\item In order to obtain the effort that is only due to the deployment environment, the work to perform any customization was excluded. This corresponds to the cost of automation, the last component of $C_i$ listed earlier in this section.
\item The service provider who is responsible for the deployment can have its own assets (tools, methodology, tacit or documented knowledge, skilled personnel) which can significantly impact the deployment and sustenance effort. We quantified this as a unitary metric called ``asset leverage factor'' $x, 0\leq x\leq1$ (smaller $x$ means that the service provider has better assets to reduce the deployment effort) and divided the observed effort with this leverage factor. We did so, in order to obtain the effort number that is only dependent only on the quadrant and the nature of the industry, without any influence of the skill that the service provider possesses. The notion of this factor $x$ will be described later in this paper.

\end{itemize}

\end{itemize}
The results of our experiment are summarized in Table~\ref{tab:data4deltaw} and graphically plotted in Figure~\ref{fig:delta-graph}. The X-axis of Figure~\ref{fig:delta-graph} marks the number of workloads to be deployed on the Quadrant \#1 cloud for the chosen deals. The normalized effort estimate to deploy the hosting cloud and to sustain it for one year (in person months) forms the Y-axis. As can be seen, the effort estimate tends to an industry-dependent plateau beyond a particular number of workloads, $W_i$. \footnote{Though we focus on Q1 here, we also repeated this experiment for other quadrants and obtained similar results. We don't include such details for brevity.} This value of $W_i$ was recorded as the "CLIC Constant" or $\delta_w$ for that industry. The CLIC Constant ($\delta_w$) thus decides the rate of growth of the hybrid complexity inside any quadrant of the cloud deployment graph depicted in Figure~\ref{fig:clic}. As the number of workload components in a quadrant increases beyond the threshold dictated by $\delta_w$, the effort needed to deploy and sustain the portion of the hybrid cloud deployment associated with that quadrant asymptotes to the maxima for that quadrant

Practitioners can repeat this experiment for other industries and decipher the respective $\delta_w$ based on the dominant shape of the curve for the industry in question.

To understand the implications of $\delta_w$, let us look at the graph plots for the health care and airline industries in Figure~\ref{fig:delta-graph}. As can be seen, the $\delta_w$ for the health care industry is observed as being less than that for the airline industry. This messages the fact that the cloud stack for the health care is generally more complex to build than that for the latter, all other aspects (such as the number of workloads, their inter-quadrant distribution and custom integration requirements) remaining the same. Similarly, the $\delta_w$ for the finance industry is lesser than the $\delta_w$ for the retail industry. This means that even with a smaller number of workloads on either side of the CLIC line, the hybrid complexity of a banking solution can potentially be higher than a retail solution that has a relatively larger number of workloads scattered across the four quadrants of Figure~\ref{fig:clic}. 

\begin{table} [ht]
\centering
\caption{Empirical Observations to Determine $\delta_w$} %\centering
\tabsize
%\begin{tabular}{|c|p{0.7in}|c|c|p{2.1in}|p{0.9in}|} 
\begin{tabular}{|c|p{0.7in}|c|p{1.4in}|p{0.3in}|} 
\toprule
\# & Industry &\#Deployments & Normalized Effort & $\delta_w$ \\
\midrule
1 & Finance & 15 & Finance curve, Fig~\ref{fig:delta-graph} & 6 \\ 
2 & Healthcare & 15 & Healthcare curve,Fig~\ref{fig:delta-graph} & 8 \\ 
3 & Retail & 15 & Retail curve, Fig~\ref{fig:delta-graph} & 10 \\ 
4 & Airline & 15 & Airline curve, Fig~\ref{fig:delta-graph} & 15 \\
5 & Manufacturing & 15 & Manufacturing curve, Fig~\ref{fig:delta-graph} & 10 \\ 
6 & Telecom & 15 & Telecom curve, Fig~\ref{fig:delta-graph} & 6 \\
\bottomrule
\end{tabular}
\label{tab:data4deltaw}
\vspace{-5pt}
\end{table}
The results from the analysis performed in this section gives rise to the following four properties that we will use to build a model for measuring hybrid complexity in the next section:
\begin{enumerate}[noitemsep,topsep=0pt,parsep=0pt,partopsep=0pt]
\item The contribution of each side of the CLIC (Figure~\ref{fig:clic}) to the hybrid
  complexity of the software system in question is approximately equal.
\item The deployment costs (and hence the complexity) has a 2:5:1:2 ratio across the four quadrants $Q_1$:$Q_2$:$Q_3$:$Q_4$. 
\item The rate of growth of deployment costs (and hence the complexity) asymptotes to a certain value as the number of workload components in the associated quadrant increase beyond the threshold dictated by $\delta_w$. 
\item The aforementioned rate of growth and threshold depends on the industry or the sector that the workload belongs to, which can be empirically determined.

\end{enumerate}

\subsubsection{Hybrid Complexity $H(w)$\\}

We now build a model to quantify hybrid complexity.\\
In Figure~\ref{fig:clic}, let $W_{1}\cdots W_{4}$ be the number of workload components in quadrants \#1 to \#4 respectively. The number of workload components to the right of CLIC = $W_{i>CLIC}=W_2$ and the number of workload components to the left of CLIC =
$W_{i<CLIC}=W_{1}+W_{3}+W_{4}$. 

This inter-quadrant distribution of $W_i$ aims to minimize the total deployment and management cost

where $C_i$ is the deployment cost of the workloads in Quadrant \#i as labeled in Figure~\ref{fig:clic}. We define the degree of hybrid complexity of the ideal deployment architecture (one that optimizes cost $C$) as follows:

\[H(w)=\frac{W_{i>CLIC}}{2\times W_{i>CLIC}+\delta_w}+\frac{W_{1}}{5\times W_{1}+\delta_w}+\frac{W_{3}}{10\times W_{3}+\delta_w}+\frac{W_{4}}{5\times W_{4}+\delta_w}
;~~~0\leq H(w)\leq1~~~~~~~~~~~~[1]
\]

where $\delta_w$ is the {\em CLIC Constant} introduced in the previous section, which depends on the class of
the workload such as its industrial grouping. The $H(w)$ of any software system lies between 0 and 1. The larger the value of $H(w)$, the more hybrid is the deployment architecture of workload $w$ and more complex is the ensuing deployment. Recall that our empirical study placed the relative costs across the four quadrants in the ratio $2:5:1:2$. This was used to arrive at the constants in the denominators of Equation 1; for example, the constant in the denominator of the third term is $10$ because it should asymptote to $\frac{1}{(2+5+1+2)}$ or $0.1$ as the number of workloads in Quadrant \#3 increase beyond the threshold dictated by $\delta_w$. Similarly, the constant in the denominator of the fourth term is $5$ because it should tend to $\frac{2}{2+5+1+2}$ or $0.2$ as the number of workloads in Quadrant \#4 increase beyond $\delta_w$.\\

The assignment of workloads along the four quadrants of
Figure~\ref{fig:clic} does not necessarily remain constant; it could change as
the attributes of the workload reshape over time. It is possible that
the demands on isolation and control are service-driven or
business-driven dynamically. For instance, consider a workload
pertaining to incident ticket management. As the number of tickets of
high severity decrease below a threshold for a certain period of
time, one may decide to relinquish some notches from the existing level of architectural control. 
Consequently, this workload may be moved to a public cloud from a private cloud. An example of a
business-driven scenario is the willingness to relax isolation
requirements on a system during time intervals when commercial
activity wanes, for example, during the first quarter of the year for
retail vendors. In short, workloads could move from one quadrant of
Figure~\ref{fig:clic} to another as a function of time. Thus Equation
1 can be more generally expressed as follows:
\[H(w,t)=\frac{W_{(i>CLIC,t)}}{2\times W_{(i>CLIC,t)}+\delta_w}+\frac{W_{1,t}}{5\times W_{1, t}+\delta_w}+\frac{W_{3,t}}{10\times W_{3,t}+\delta_w}+\frac{W_{4,t}}{5\times W_{4, t}+\delta_w}~~~~~~~~~~~~[2]\]
We will demonstrate the application of both Equation 1 and Equation 2
using case studies in Section~\ref{sec:casestudy}.

A note on the CLIC Constant before we end this subsection. One aspect that has to be taken into account while determining $\delta_w$ 
is that industries often intersect. An e-commerce vendor might generate 
revenue via the financing route. Customers would potentially be allowed 
to purchase on credit and maintain running accounts on which interest is 
levied. Such a retail enterprise, thus, has workloads both in the retail sector 
and in the finance sector. The value of $\delta_w$, thus, needs to be chosen
accordingly to the workload category rather than the industrial
classification of the enterprise. The retail and financial workload deployments 
in this example would thus be treated as separate hybrid environments with 
different associated $\delta_w$s. 
The Government is another sector
than commonly spans industries. Government IT needs to serve
departments touching the breadth of the economic spectrum ranging from
finance and health, to telecom and manufacturing. The value of
$\delta_w$, thus, has to be chosen appropriately based on the
workload\textquotesingle s industrial grouping rather than the
enterprise\textquotesingle s official industry classification.
\subsection{Predicting Hybrid Deployment \& Sustenance Effort}
\label{sec:predicting-effort}
Deploying and sustaining a hybrid cloud is a complex and
time-consuming activity. Consequently, the approach to estimate the
hybrid cloud deployment and management effort has largely been
ad-hoc. We have observed that the hybrid complexity measure $H(w)$ can
play a significant role in improving the estimation of deployment and
management effort. We attempt to model the deployment and
management effort $\widehat E(w)$ of a set of workloads $w$ constituting a software system as:
\[\widehat E(w) = H(w)\times K \times \frac {x}{y}~~~~~~~~~~~~~~~~~~~[3] \]
$\widehat{E}(w)$ in Equation 3, in effect, is influenced by the following:
\begin{enumerate}[noitemsep,topsep=0pt,parsep=0pt,partopsep=0pt]
\item A constant $K$, referred to as the {\em complexity-to-effort
  constant}, whose value is dependent on the hybrid cloud deployment
  service provider, and is determined based on prior experience, given
  the available level of skill and expertise of the service provider.
\item An {\em asset leverage factor} $x$ ($0<x<1$), whose value depends on the
  quality of internally available assets and tools that can be used to
  assist the service provider to automate the deployment.
  The smaller the value of $x$, the more sophisticated is the assistant toolset to automate the hybrid cloud deployment in  question.
\item A {\em custom work complexity factor} $y$ ($0<y<1$), whose value
  depends on custom work that needs to be implemented around the
  hybrid deployment. This includes, for example, bespoke workflows to
  integrate the hybrid deployment with service management tools, or
  custom code development to fuse hybrid cloud authentication with
  pre-existing access directories. Lesser the amount of required
  custom work, nearer $y$ is to 1.
\end{enumerate}

$H(w)$ is a necessary, but not solely sufficient component to
calculate $\widehat E(w)$ as shown in Equation 3. However, $H(w)$ is the sole
provider-independent component in the equation. The rest are either
service provider-specific ($K$ and $x$) or depend on custom requirements of 
the hybrid cloud that is being implemented ($y$).

To calculate $K$ for the service provider applicable to our case, we randomly chose a set of hybrid cloud solutions deployed over the past one year in the organization. For each deployment, we computed its $H(w)$ using Equation 1, and observed the actual effort $E(w)$ expended to setup and sustain the solution for one year. We then estimated the asset maturity within the organization for the industry in question ($x$) and the custom effort required for the chosen deal ($y$). Next we determined the value of $K=\frac{E(w)}{H(w)}*\frac{y}{x}$ for the deployment. We computed the overall $K$ for the service provider as the cumulative moving average of $K$ across the set of deployments. The result of the calculation is shown in Table ~\ref{tab:subdata2verifymodel}. 

It is prudent to periodically fine-tune the value of $K$ as past deployment data accumulates, so that currency is maintained on the epoch over which the cumulative moving average is being captured. This is because the value of $K$ is a function of time and can change as internal teams and organizational processes mature. Similarly, it is necessary to re-calibrate $x$ on a timely basis since it can change as industry-specific reusable assets get built.

In Section~\ref{sec:casestudy}, we will present data from twelve client case studies to evaluate the effort estimate model proposed in Equation 3. The variance of observed results ($E(w)$) from what Equation 3 predicts ($\widehat E(w)$) is presented in Table~\ref{tab:data2verifymodel} of Section~\ref{sec:eval-model}.

\vspace{-5pt}
\subsection{Methodology to Converge on the Effort Estimate of Hybrid Deployments}

We end this section by summarizing the methodology that we propose to converge on the effort estimate of the ideal hybrid
deployment of a software system:

\begin{enumerate}[noitemsep,topsep=1pt,parsep=0pt,partopsep=1pt]
\item Demarcate the set of workloads that require high levels of
  isolation and high levels of control; in other words, draw the CLIC
  contour that determines the layout of the four quadrants in the
  cloud deployment graph. Count the number of workloads to the right
  hand side of the CLIC and compute $W_{Q2}$.
\item For the workloads to the left hand side of the CLIC, further assess and
  separate them based on isolation and control requirements, 
  into $W_{Q1}$, $W_{Q3}$ and $W_{Q4}$.
\item Identify the industrial grouping of the workload that
  empirically determines the rate of growth of hybrid complexity
  as a function of the number of workloads, and ascertain the
  applicable $\delta_w$ (see Figure~\ref{fig:delta-graph}).
\item Apply the values determined above to Equation 2 and deduce the
  hybrid complexity ($H(w)$) of deploying this software
  system on the best-fit cloud topology.
\item Apply the $H(w)$ determined above to Equation 3 to predict the effort 
 estimate needed to deploy and sustain the associated hybrid cloud environment.
\end{enumerate}

Discovering $H(w)$ (and the associated $\widehat E(w)$) to deploy and run a software system 
on the best-fit hybrid cloud brings the following benefits:

\begin{enumerate}[noitemsep,topsep=1pt,parsep=0pt,partopsep=1pt]
\item It provides a high level view of the relative work and cost
  estimate to deploy these workloads, and manage them during steady
  state operations. This in turn, yields a measure of the TCO
  (total cost of ownership) of the hybrid deployment in question.
\item It can function as a tool to compare the relative hybrid
  deployment intricacies of two or more software systems.
\item Designing a hybrid deployment is generally dependent on numerous
  variables, such as performance SLAs, adherence to a set of
  regulatory compliance requirements, degree of tolerance to workload collocation,
  as well as the perception of diminished security in off-premise
  systems. $H(w)$ encapsulates all of these complexities latent in a
  hybrid deployment in the form of a single number. The ability to
  quantify the complexity in this manner allows for easier business
  and architectural decisions.
\end{enumerate}

\vspace{-12pt}
\section{Evaluating the Hybrid Deployment Model: Application to Industrial Case Studies}
\label{sec:casestudy}
In this section, we will evaluate the hybrid deployment model that we constructed above.
We took three real life client examples from four industries (for a total of twelve client deals)
where we had designed hybrid cloud deployment topologies. None of these client solutions were part of the data set that was leveraged to construct our model. We applied our model to predict the hybrid cloud deployment effort $\widehat E$ in each case and also observed the actual work effort $E$ that was incurred. The results and the ensuing variance are presented at the end of this section in Table~\ref{tab:data2verifymodel}. For ease of presentation, we chose and expanded one client example from each industry for a total of four case studies in this section, but the results for all twelve deals that we observed are presented in Table~\ref{tab:data2verifymodel}.

Our first case study is a software system from the retail industry,
the second from financial services, the third from the health care
sector, and the fourth from the airline industry. In each case, 
we show how we performed a workload analysis,
drew the CLIC line that helped derive the hybrid deployment
architecture, derived the hybrid deployment complexity $H(w)$, 
and observed predicted $\widehat E(w)$ and actual $E(w)$. We
also present our observations regarding the CLIC constant $\delta_w$.

In order to explain the deployment scenario, we
consider an architecture reference model of the industry
associated with each example. A reference model for an industry
provides sufficient information about the functional entities and
their relationships in the given industry context, while remaining
abstract so that it does not delve into non-essential details of
specific systems belonging to the industry.
\subsection{Retail Sector}
\label{sec:retail-casestudy}
\begin{figure} 
\centering 
\includegraphics[scale=0.45]{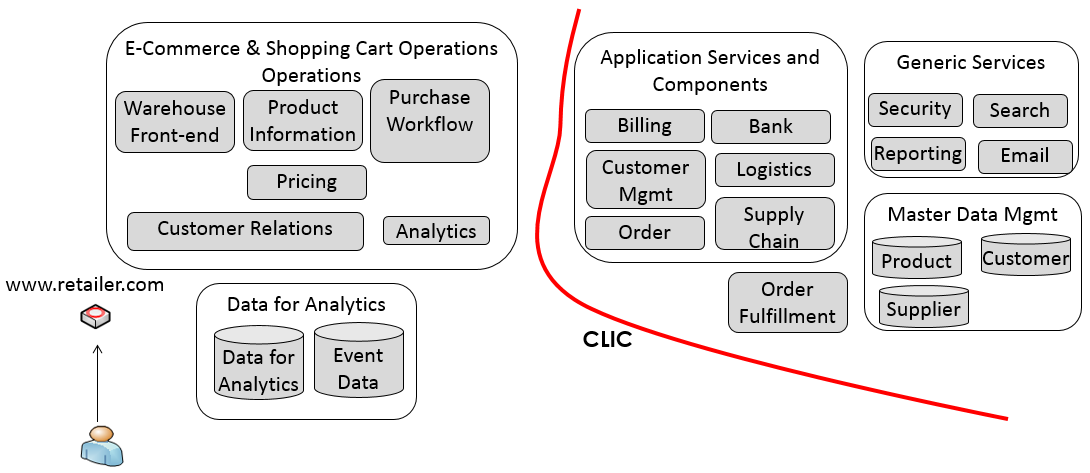}
\caption{A Simplified Architecture Reference Model for Retail}
\label{fig:retail}
\vspace{-10pt}
\end{figure}

Let us consider a typical online retail system. We have considered a
simplified version of the reference model described
in~\cite{Aulkemeier2016} and depicted that in Figure~\ref{fig:retail}.
Retail software can typically be classified into two main categories~\cite{Aulkemeier2016}:
\begin{enumerate}
[noitemsep,topsep=1pt,parsep=1pt,partopsep=1pt]
\item Front-end transactions before potential end-customers populate
  an item into the shopping cart. This includes the e-commerce 
  portal, operations on the retailer\textquotesingle s catalogue, and 
  analytics to derive purchase recommendations. We call these
  "above the shopping cart" processing.

\item Back-end processing after an item has been added to the shopping
  cart. This includes order management, billing, and 
    fulfillment. We call these "below the shopping cart" processing.
\end{enumerate}

Processing "above the shopping cart" has to be production-grade, but
action "below the shopping cart" is mission critical. The former
includes the stage when customers browse and search the
retailer\textquotesingle s catalogue, evaluate personalized
recommendations, and manage accounts. The latter commences after the
customer checks out the purchase, and encompasses payment and delivery. 

Big retailers are reluctant to move "below the cart" processing
off-premise to a public cloud because this portion is built in a
highly redundant fashion with clustered dedicated components without
single points of failure. It is difficult to realize such deployment
architectures on many public clouds, but possible on private clouds
where the retailer can exercise high degree of architectural
control. For example, a clustering solution that needs a dedicated
network link to carry heartbeats is hard to implement on a
multi-tenant cloud. Or if workloads are sensitive to regulation 
-– Payment Card Industry (PCI) compliance in this case -– it is 
easier to host them on a private cloud over which the retailer can 
impose high levels of isolation. Retail systems also maintain personal 
data such as customer credit card information and shipping addresses, 
which are sought to be stored as close to on-premise as possible.
Most retailers are, however, eager to move functionality existing
”above the cart” to public clouds. In addition to being less mission
critical, these functions tend to be more customer focused. For
instance, a typical retail business provides users a personalized
shopping experience, typically by leveraging social media and mobile
devices. Such system components are comfortable living
off-premise. Additionally, modern retail systems have a set of
constituents that generate data-driven insights such as sales
forecasts, type and timing of promotions, and pricing
strategies. These systems also thrive on public cloud based
infrastructure, where they heavily use MapReduce algorithms and
leverage a service oriented design where application services are
exposed and consumed via published APIs.

\begin{figure} 
\centering 
\includegraphics[scale=0.45]{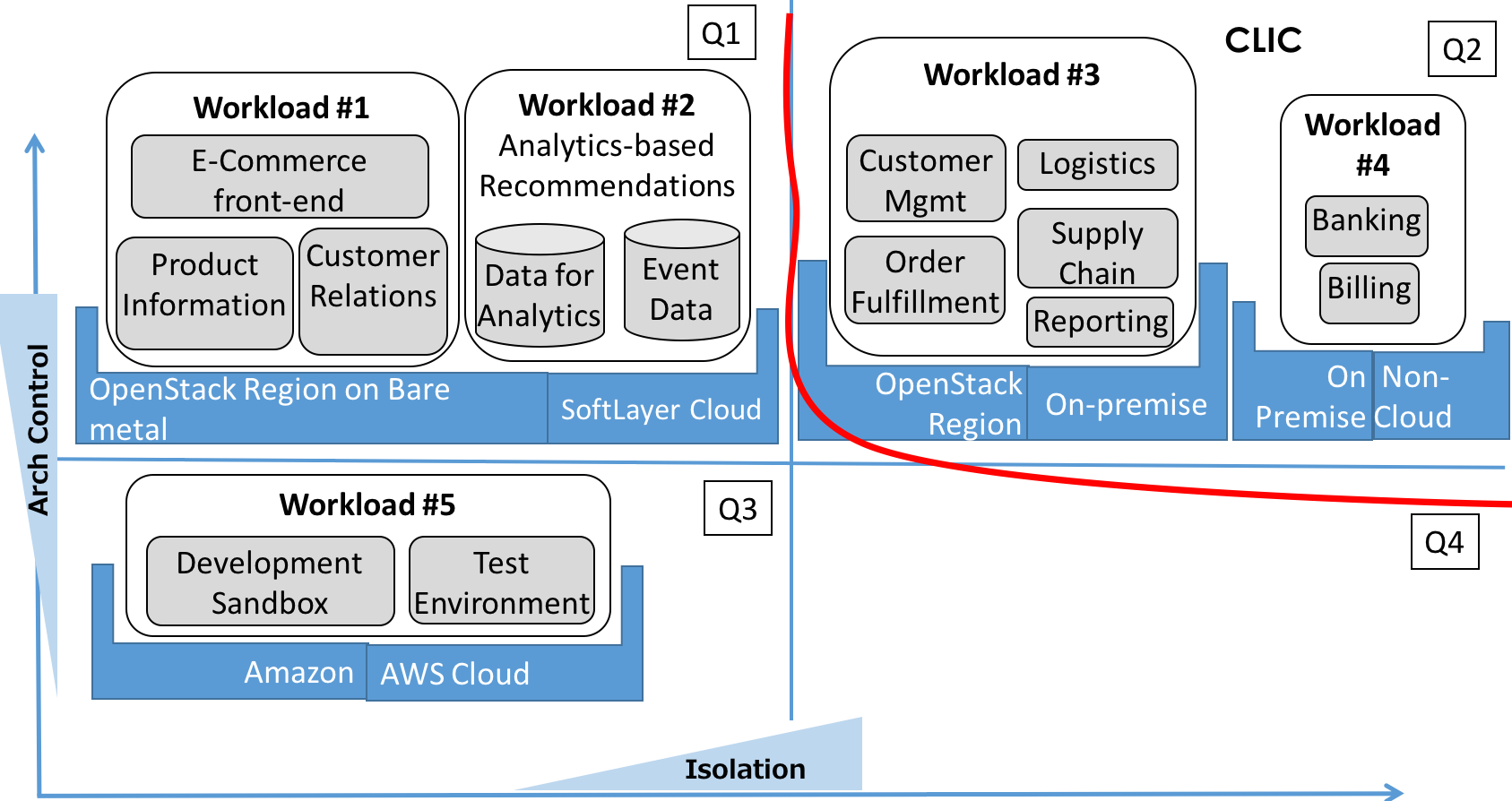}
\caption{H(w, t=Apr-Dec)=0.29 for a Retail Software System}
\label{fig:retail_dec}
%\vspace{-10pt}
\end{figure}

Figure~\ref{fig:retail_dec} depicts a deployment of our retailer case
study. The workloads generally conform to, but is a subset of the
reference model shown in Figure~\ref{fig:retail}. Workloads to the
right of the CLIC (the ones that fall in Quadrant \#2 of the
diagram depicted in Figure~\ref{fig:clic}) comprise mainly of "below
the shopping cart" software). The deployment options for these
workloads are permutations of what is articulated in the box
superimposed on Quadrant \#2 of Figure~\ref{fig:clic}. In the context
of this customer, the choice was a combination of the following:
\begin{enumerate}
[noitemsep,topsep=0pt,parsep=0pt,partopsep=0pt]
\item A traditional non-cloud infrastructure to hold legacy backend
  retail applications running on mainframes. These were the
  applications that enabled financial transactions and billing.
\item An OpenStack cloud region~\cite{openstack} deployed on-premise
  on the client\textquotesingle s data center to host order fulfillment
  applications deployed atop an Oracle Real Application Cluster (RAC)
  for high availability and redundancy.
\end{enumerate}

Workloads to the left of the CLIC in Figure~\ref{fig:retail_dec} are
spread across Quadrants \#1, and \#3. The main workload in this 
category was the e-commerce frontend and a recommendations engine 
built around a Hadoop Big Data framework. On the basis of the waxing 
and waning of commercial activity over the past several years, this 
retailer concluded that during the busy months of the year they needed 
architectural control at the hypervisor layer to stitch together a high 
availability (HA) model that provided a low Mean Time to Recover (MTTR) 
and high Mean Time to Failure (MTTF). This was realized by deploying a 
dedicated OpenStack cloud region on top of virtualized bare metal servers 
on the IBM SoftLayer Public Cloud~\cite{softlayer} as depicted in 
Figure~\ref{fig:retail_dec}.

\begin{figure} 
\centering 
\includegraphics[scale=0.45]{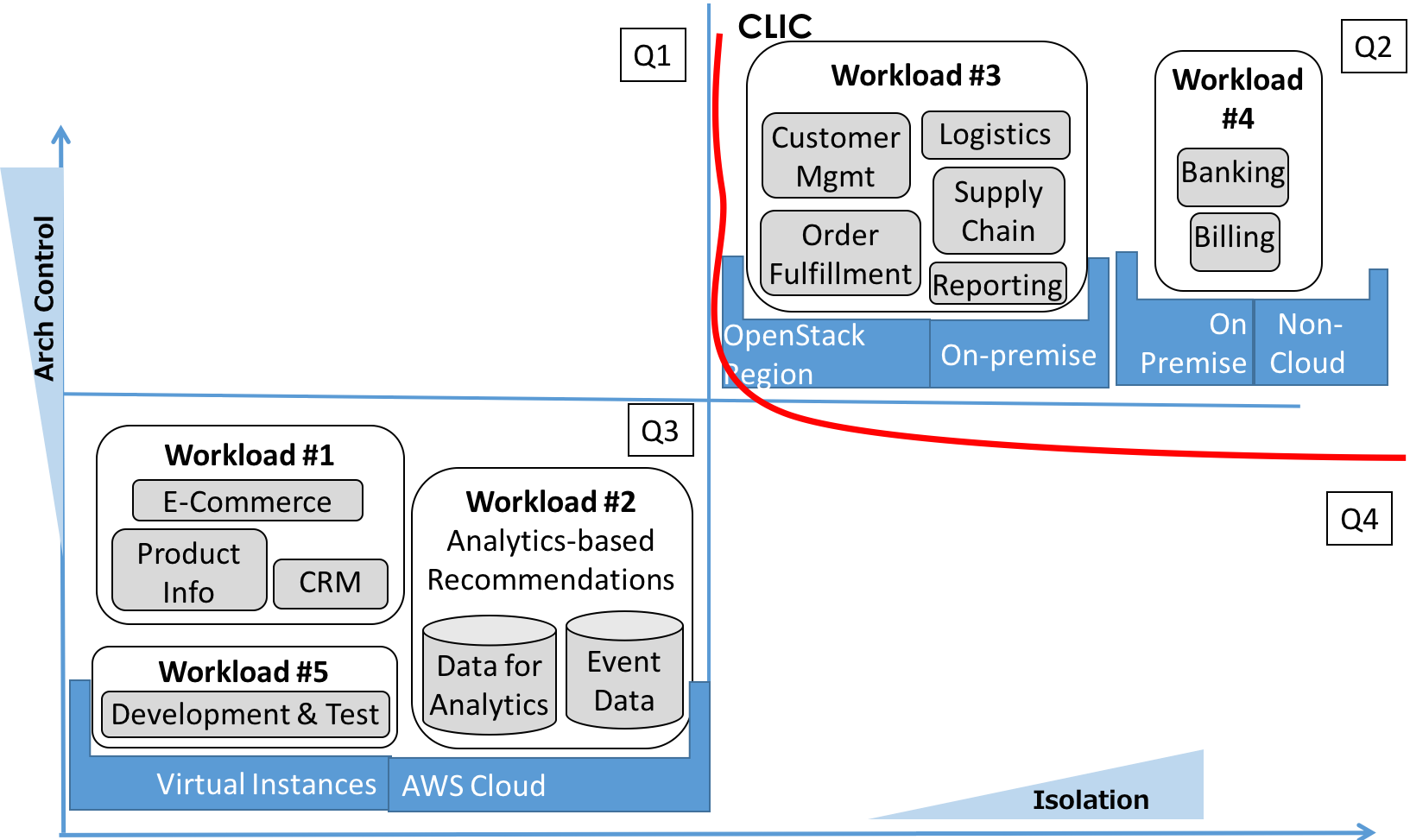}
\caption{H(w, t=Jan-Mar)=0.22 for a Retail Software System}
\label{fig:retail_mar}
\vspace{-10pt}
\end{figure}

During the lean months of the year (primarily the first quarter from
January to March), however, availability requirements could be
relaxed, and consequently, hypervisor-level control was not
needed. During this period, following Equation 2, workloads residing
in Quadrant \#1 of Figure~\ref{fig:clic}, can be allowed to move down
to Quadrant \#3. This resulted in a simplified hosting model of wherein
the workloads could be hosted on public cloud virtual instances on
Amazon Web Services (AWS)~\cite{amazoncloud}. This scenario is shown in Figure~\ref{fig:retail_mar}. 
Note that there is inter-quadrant movement in this case, but not across the CLIC line.

Development and test workloads reside in Quadrant \#3 throughout the
year, hosted on AWS virtual machines [5]. The customer, thus, had four
kinds of workloads as depicted in Figure 3, spread across three
quadrants. We applied Equation 2 to measure the hybrid deployment
complexity of this software system as a function of time. Note that we
empirically observed $\delta_w$ for retail workloads as $10$ in 
the industry plot in Figure~\ref{fig:delta-graph}.

For the partitioned deployment of Figure~\ref{fig:retail_dec}, we get,
$H(w,\textrm{Apr-Dec})=\frac{2}{14}+\frac{2}{20}+0+\frac{1}{20}=0.29$.

For the partitioned deployment of Figure~\ref{fig:retail_mar}, we get,
$H(w,\textrm{Jan-Mar})=\frac{2}{14}+0+0+\frac{3}{40}=0.22$.

These numbers quantify the hybrid deployment complexity that we
predict for this retail business. Note that if there are no workloads 
in Quadrant $i$ for a hybrid deployment, the corresponding term in 
Equation 1 (and Equation 2) becomes 0. Table~\ref{tab:data2verifymodel} 
translates the calculated hybrid complexity to cost in terms of person month 
effort estimates; it also provides conclusions pertaining to the 
complexity of this retail case study relative to other case studies 
that we discuss in this paper.
\subsection{Banking Sector}
\begin{figure} 
\centering 
\includegraphics[scale=0.5]{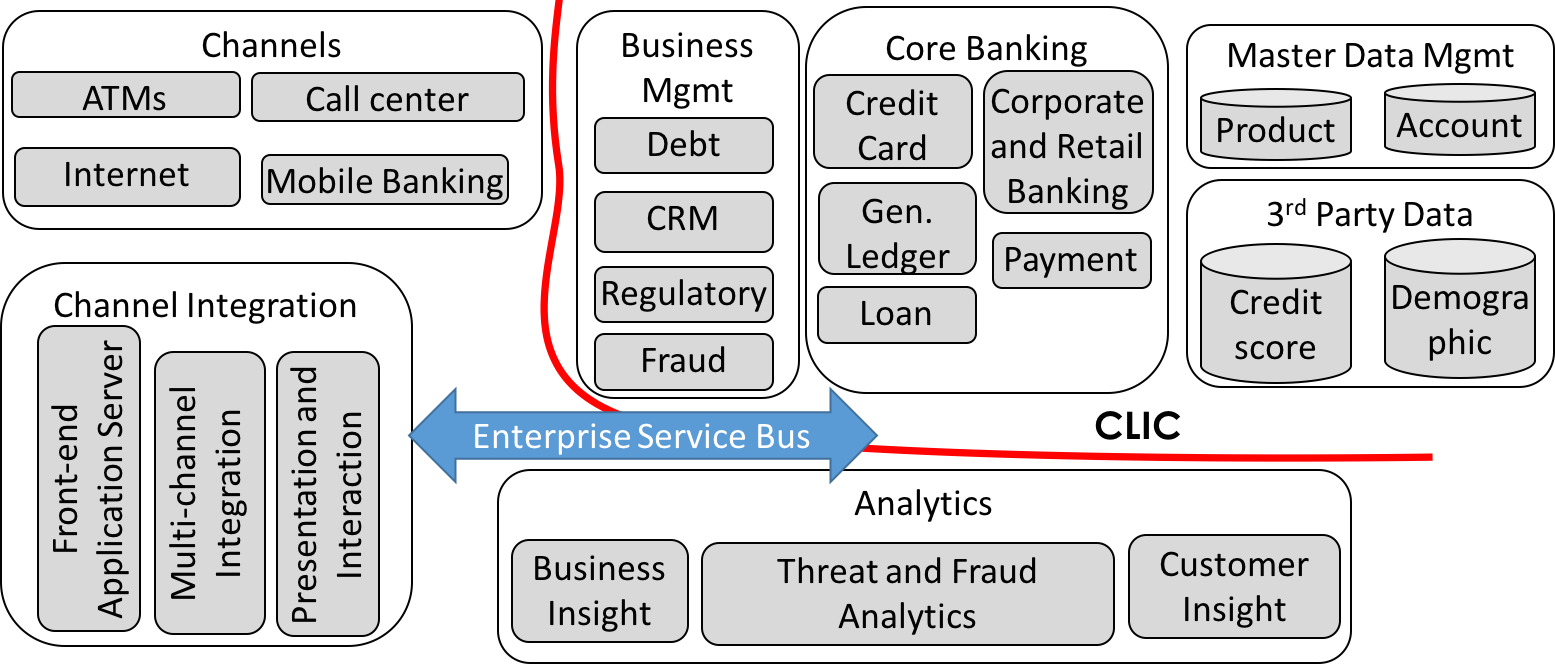}
\caption{Simplified Architecture Reference Model for Bank}
\label{fig:bank-clic}
\end{figure}

\begin{figure} 
\centering 
\includegraphics[scale=0.46]{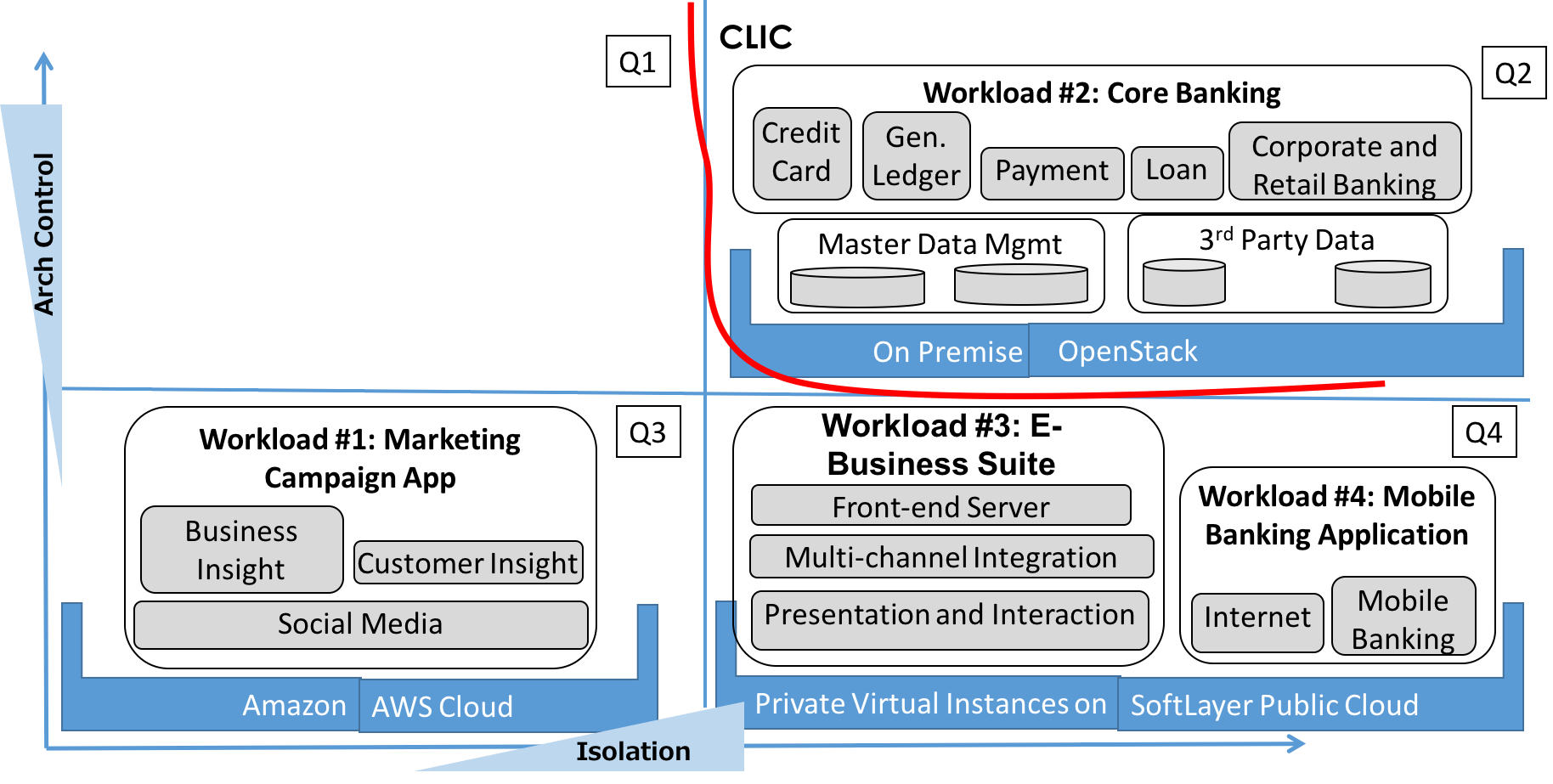}
\caption{H(w)=0.31 for a Banking Software System}
\label{fig:bank-graph}
%\vspace{-5pt}
\end{figure}

We observed that the financial industry has a natural affinity to
hybrid clouds. Enterprises prefer mission critical and regulation
sensitive workloads to run on private clouds where the business can
exercise architectural control. Less critical production workloads are
migrated to public clouds. Let us consider a generic banking sector
reference architecture shown in Figure~\ref{fig:bank-clic}, which is a
simplified version of the banking reference model described
in~\cite{Keen2009}. Further, consider Figure~\ref{fig:bank-graph},
which depicts the infrastructure that applies to our case
study. Figure~\ref{fig:bank-graph} conforms to, but is a subset of,
the reference model shown in Figure~\ref{fig:bank-clic}, and explains
the hybrid cloud deployment pattern that we designed to host a
bank\textquotesingle s software system.
There are four workloads in this case comprising of three sets:
\begin{enumerate}
[noitemsep,topsep=0pt,parsep=0pt,partopsep=0pt]
    \item Core banking applications are mission critical. Even a
      few minutes of downtime of these workloads can translate to
      millions of dollars of revenue and brand image loss. These
      applications transact with customer financial
      records. Regulatory requirements demand isolation for this
      ensemble of workloads; custom security requirements call for
      architectural control as well. This portion of the
      bank\textquotesingle s software, thus, firmly resides in
      Quadrant \#4. These workloads were
      hosted on an OpenStack-based private cloud~\cite{openstack}
      implemented on-premise as shown in Figure~\ref{fig:bank-graph}.
    \item Two workloads that reside in Quadrant \#4 on the left hand
      side of the CLIC: An e-business suite centered round the
      bank\textquotesingle s web portal, and an application that
      generates mobile alerts based on state changes reported by the
      core banking environment. These software components were hosted
      on virtual private instances on the IBM SoftLayer Public
      Cloud~\cite{softlayer}. As discussed earlier, virtual private
      instances offer server level isolation for a customer.
    \item A marketing campaign application that draws insight from
      social media interactions. Because the analytics engine associated with
      this application was built around a Hadoop Big Data engine,
      virtual instances on a public cloud offered the most advantageous
      deployment infrastructure for this application.
\end{enumerate}

The overall deployment architecture for the bank\textquotesingle s
software system is depicted in Figure~\ref{fig:bank-graph}. The hybrid
complexity was calculated by the application of Equation 1. Note
that we have empirically observed $\delta_w$ for financial workloads as $6$
in the industry plot in Figure~\ref{fig:delta-graph}.
\[H(w)=\frac{1}{8}+0+\frac{2}{16}+\frac{1}{16}=0.31\].
\vspace{-6pt}
\subsection{Health care Sector}
\begin{figure} 
\centering 
\includegraphics[scale=0.55]{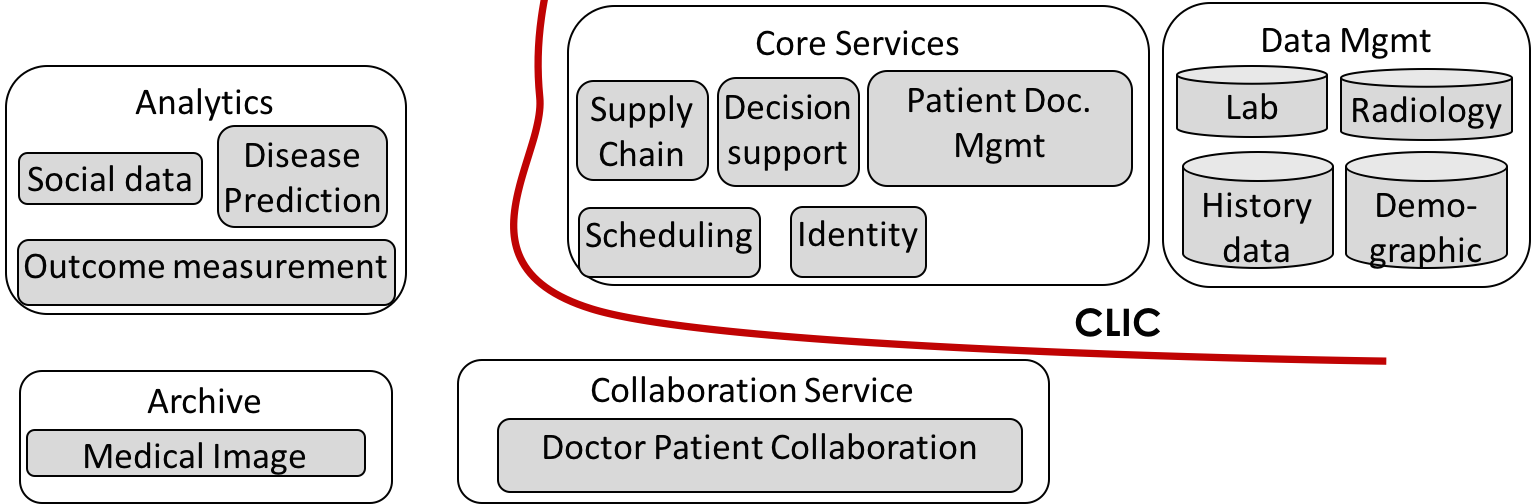}
\caption{Simplified Architecture Reference Model for Health Care}
\label{fig:healthcare-clic}
\vspace{-10pt}
\end{figure}

\begin{figure} 
\centering 
\includegraphics[scale=0.46]{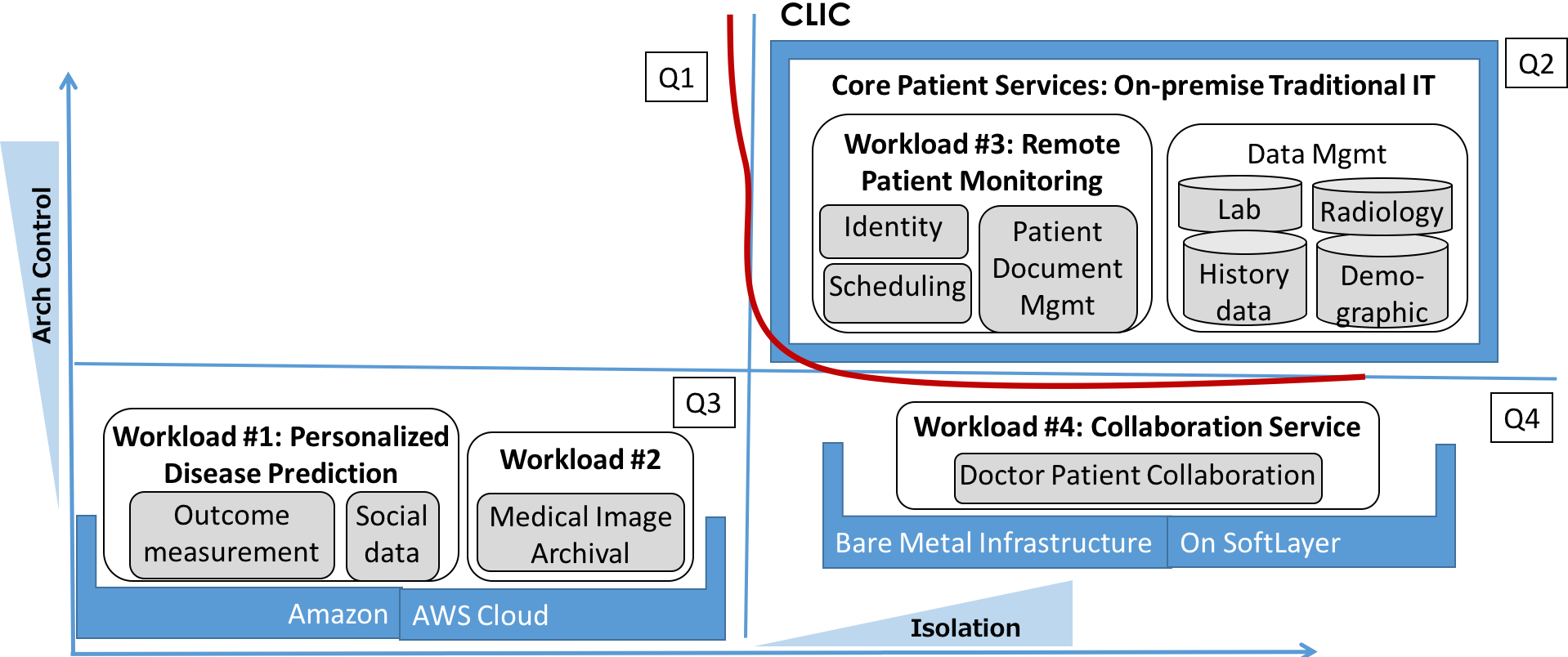}
\caption{H(w)=0.25 for a Health Care Software System}
\label{fig:healthcare-graph}
\vspace{-10pt}
\end{figure}

Health care is being transformed by technology, cloud, and the
Internet. The use cases are many and varied: collaborative care
without spatial boundaries, remote patient monitoring, medical
imaging, mobile-enabled hospital management systems, and innovative
and predictive Systems of Insight. Figure~\ref{fig:healthcare-clic}
shows the hybrid cloud deployment pattern relevant for a modern health
care service. The architecture reference model shown in this figure
has been adapted from~\cite{Orvis2008}. The case study in question has
workloads that constitute a subset of the aforementioned reference
model and is depicted in Figure \ref{fig:healthcare-graph}.
The client case study in question had four workloads:

\begin{enumerate}
[noitemsep,topsep=0pt,parsep=0pt,partopsep=0pt]
    \item A software system that queried and monitored the
      patient\textquotesingle s cardiac parameters. This software was
      an FDA Class-3\footnote{Food and Drug Administration (FDA) is
        the agency that regulates drugs and medical devices in the
        United States. A device is classified as Class-3 by the FDA if
        its failure can be life threatening; hence devices under this
        category are subject to the highest level of regulation.} system, thus
      sensitive to regulation. This system falls on the right side of
      the CLIC and was hosted on a traditional non-cloud IT
      infrastructure.
    \item A software suite built around analytics and social media to
      predict and prevent diseases based on past diagnostic data,
      while also aiming to improve outcomes by measuring clinical
      results on patients. The framework on which this workload
      depended, lent itself to being a natural fit for Quadrant \#3. 
      This part of the overall
      workload was hosted on virtual instances on the Amazon
      Cloud~\cite{amazoncloud}.
    \item A collaboration system to remotely connect doctors and
      patients. Given that the stakeholders would be geographically
      separated, the communication needed to be rapid and secure. The
      latter brought in requirements on isolation, best satisfied by
      deployment options associated with Quadrant \#4. 
      This workload, thus, was hosted on single-tenant bare metal servers 
      on the IBM SoftLayer Public Cloud~\cite{softlayer}.
    \item An archival system for storing medical images. This data was
      infrequently accessed and could tolerate large retrieval access
      times. An off-site storage that was a replacement for tape drives
      was the need, and hence this workload also falls on Quadrant
      \#3. The storage solution chosen in this case was Amazon
      Glacier services~\cite{amazonglacier}.
\end{enumerate}

The overall deployment architecture for the health care
provider\textquotesingle s software system is depicted in
Figure~\ref{fig:healthcare-graph}. The hybrid complexity was
calculated by the application of Equation 1. Note that we have
empirically observed $\delta_w$ for health care workloads as $8$ 
in the industry plot in Figure~\ref{fig:delta-graph}.
\[H(w)=\frac{1}{10}+0+\frac{1}{13}+\frac{2}{28}=0.25\].
\vspace{-6pt}
\subsection{Airline Sector}
\begin{figure} 
\centering 
\includegraphics[scale=0.4]{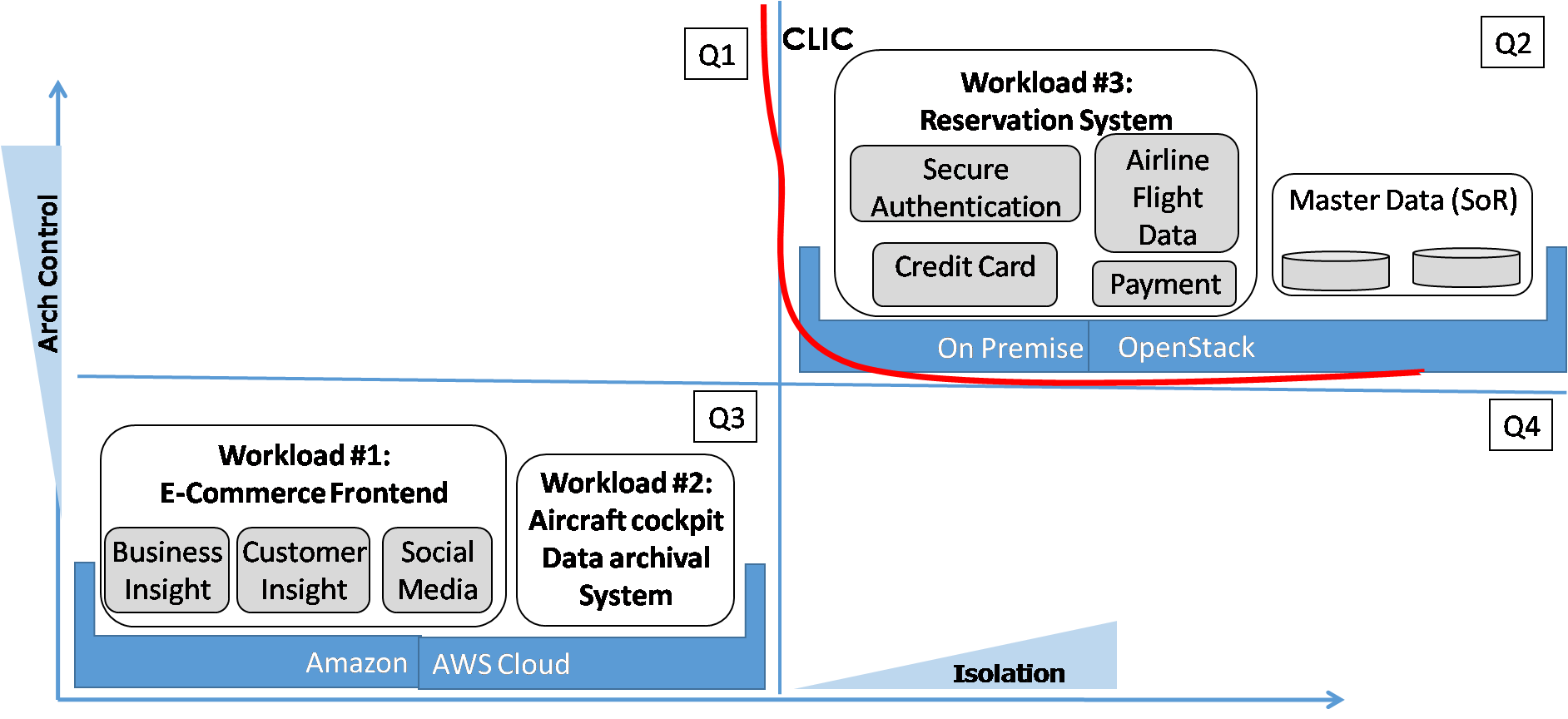}
\caption{H(w)=0.12 for an Airline Software System}
\label{fig:airline}
\vspace{-2pt}
\end{figure}

Our fourth case study is from the aviation industry, which is part of the
larger transportation sector. Figure~\ref{fig:airline} describes a customer
scenario where a mission critical airline reservation system that
falls on the right side of the CLIC was hosted on a secure dedicated
on-premise private cloud. The customer-engaging front-end portal that belonged to
the left of the CLIC was deployed on the Amazon Public
Cloud~\cite{amazoncloud}.

The hybrid complexity was calculated by the application of Equation 1. 
Note that we have empirically observed $\delta_w$ for the Airline sector as $15$
in the industry plot in Figure~\ref{fig:delta-graph}.
\[H(w)=\frac{1}{17}+0+0+\frac{2}{35}=0.12\]

\subsection{Evaluating the Hybrid Deployment Model with Results from the Case Studies}
\label{sec:eval-model}
In this section, we use data from our case studies to evaluate 
our model expressed via Equation 1, Equation 2 and Equation 3. We apply the 
$H(w)$ values we calculated for our case studies to Equation 3 and record the 
effort estimate $\widehat E(w)$ that it predicts. We then compare it with the actual effort $E(w)$ that 
was expended to deploy and sustain the hybrid cloud environment associated with 
our case studies for a period of one year.

Table~\ref{tab:data2verifymodel} summarizes our findings. As mentioned earlier, we observed three application deployments from each of the four chosen industries, so the table has data points from twelve deployments. The inter-quadrant workload distribution for each client deployment is specified within brackets in column \#2. The four deployments highlighted in \textbf{bold} are the ones that we described in detail earlier in this section. All the chosen deployments were performed by one service provider organization. 

\begin{table} [ht]
\caption{Cloud Provider Specific Variables to Calculate $\widehat{E}(w)$} \centering
\tabsize
\begin{tabular}{|l|c|c|} 
\toprule
Industry & $K$ & $x$ \\
\midrule
Retail & 150 & 0.8 \\ 
Finance & 150 & 0.6 \\
Health Care & 150 & 0.8 \\
Airline & 150 & 0.7 \\
\bottomrule
\end{tabular}
\label{tab:subdata2verifymodel}
\vspace{-5pt}
\end{table}
To explain the calculations, let's consider the retail case study that we discussed in Section~\ref{sec:retail-casestudy} (row \#1 of Table~\ref{tab:data2verifymodel}). As determined in Section~\ref{sec:retail-casestudy}, this client deployment yielded an $H(w)$ of 0.29 resulting from the application of Equation 1. To apply Equation 3, we next need the two cloud service provider specific parameters, $K$ and $x$. These values are depicted in Table~\ref{tab:subdata2verifymodel}, which contains the $K$ and $x$ values for the cloud service provider organization applicable to our experiment. $K$ was empirically determined as described in Section~\ref{sec:predicting-effort}, whereas $x$ was estimated for each industry.

As can be seen from Section~\ref{sec:retail-casestudy}, the cloud service provider organization applicable to our case had an empirically determined $K$ value of $150$, 
which implies that based on internal skills  and processes of the service provider, 
a 43 person month effort can implement and sustain a hybrid deployment having an H(w) of 0.29, 
assuming lack of relevant reusable assets ($x=1$) and absence of custom requirements 
($y=1$). Our cloud service provider, however, has an asset leverage factor of $x=0.8$ for retail cloud 
deployments, which means that the availability of 
reusable assets from similar earlier deals is relatively low. The custom-multiplier $y$ for this retail deployment was 0.2, which implies that the contractual custom
requirements to make the deployment infrastructure ready for the application deployment was high. If we apply these values to Equation 3 to obtain the expected
effort estimate, we get,
$\widehat E(w)=0.29*150*\frac{0.8}{0.2}=174$~Person~Months.
We also computed the percentage variance between the predicted sizing numbers ($\widehat E(w)$) and the actual values that we observed post-deployment ($E(w)$), which in this case was $10\%$ (the last column of the first row in Table~\ref{tab:data2verifymodel}).\\ 

Our post-deployment measurements thus reveal that our method to determine the hybrid complexity of client workloads and to forecast the associated effort estimate is a reasonably accurate technique.

\begin{table} [ht]
\caption{Effort Estimate across Case Studies: Predicted versus Actual} \centering
\tabsize
\begin{tabular}{|c|l|c|c|p{1.0in}|p{0.7in}|} 
\toprule
\# & Industry & $H(w)$ & $y$ & $\widehat E(w)$ & $|\widehat E(w)-E(w)|$\\
\midrule
\textbf{1} & \textbf{Retail (Case Study \#1)} & \textbf{0.29} & \textbf{0.2} & \textbf{174 person months} & \textbf{10\%}\\ 
2 & Retail ($W_1=1, W_2=2, W_3=2, W_4=2$) & 0.38 & 1.0 & 46 person months & 4\%\\
3 & Retail ($W_1=3, W_2=3, W_3=3, W_4=2$) & 0.48 & 0.8 & 72 person months & 13\%\\
\textbf{4} & \textbf{Finance (Case Study \#2)} & \textbf{0.31} & \textbf{0.2} & \textbf{117 person months} & \textbf{15\%}\\
5 & Finance ($W_1=4, W_2=3, W_3=5, W_4=2$) & 0.62 & 0.3 & 186 person months & 15\%\\
6 & Finance ($W_1=5, W_2=2, W_3=3, W_4=5$) & 0.77 & 0.4 & 173 person months & 12\%\\
\textbf{7} & \textbf{Health Care (Case Study \#3)} & \textbf{0.25} & \textbf{0.5} & \textbf{60 person months} & \textbf{12\%}\\
8 & Health Care ($W_1=4, W_2=2, W_3=4, W_4=2$) & 0.50 & 0.6 & 100 person months & 9\%\\
9 & Health Care ($W_1=5, W_2=2, W_3=3, W_4=1$) & 0.47 & 0.3 & 188 person months & 2\%\\
\textbf{10} & \textbf{Airline (Case Study \#4)} & \textbf{0.12} & \textbf{0.3} & \textbf{18 person months} & \textbf{8\%}\\
11 & Airline ($W_1=2, W_2=1, W_3=4, W_4=0$) & 0.21 & 0.5 & 44 person months & 5\%\\
12 & Airline ($W_1=0, W_2=2, W_3=5, W_4=1$) & 0.23 & 0.7 & 35 person months & 14\%\\
\bottomrule
\end{tabular}
\label{tab:data2verifymodel}
\vspace{-5pt}
\end{table}
\begin{table} [ht]
\caption{Complexity of Non-Functional Requirements across Industries} \centering
\tabsize
\begin{tabular}{|l|p{0.35in}|p{0.49in}|p{0.45in}|p{0.37in}|p{0.54in}|p{0.55in}|p{0.55in}|p{0.35in}|}
%\begin{tabular}{|l|p{0.25in}|p{0.25in}|p{0.25in}|p{0.25in}|p{0.3in}|p{0.25in}|p{0.3in}|p{0.12in}|} %
\toprule
Industry & \# Deployments & Availability & Biz Continuity & Security & Compliance & Performance & Complexity Quotient & $\delta_w$ from Table~\ref{tab:data4deltaw} \\
\midrule
Airline        & 15 & M & M & L & L & M & 8 & 15 \\
Retail         & 15 & M & M & M & M & M & 10 & 10 \\
%Manufacturing  & 15 & L & M & H & M & M & 10 & 3 \\
Health Care    & 15 & M & M & H & H & M & 12 & 8 \\
Finance        & 15 & M & H & H & H & M & 13 & 6 \\
Manufacturing  & 15 & M & L & H & M & M & 10 & 10 \\
Telecom        & 15 & M & H & M & H & H & 13 & 6 \\
%Telecom        & 8 & L & M & H & M & M & 13 & 1 \\
%Government     & 5 & M & M & L & L & M & 13 & 1 \\ 
\bottomrule
\end{tabular}
\label{tab:data2verifydeltaw}
\vspace{-5pt}
\end{table}
\subsection{Validation of $\delta_w$}
We also performed an experiment, shown in Table~{\ref{tab:data2verifydeltaw}},
to demonstrate that our observations and conclusions on $\delta_w$ presented 
in Table~{\ref{tab:data4deltaw}} are intuitive as well. The deployments listed in Table~{\ref{tab:data2verifydeltaw}} are the same ones that we chose for our observations recorded in Table~{\ref{tab:data4deltaw}} (and hence Figure~{\ref{fig:delta-graph}}). Each row of this table grades the non-functional requirements associated with an industry, which we ascertained from actual application deployment data in that business grouping. Each cell is rated High(H), Medium(M) or Low(L) having weights of 3, 2, and 1, 
respectively, that combine to determine a \textit{complexity quotient} for that industry. Note that each entry is the average of the characteristics of all workloads in that client deployment. As seen in the table, compliance requirements imposed
on health care, finance and telecom industries are higher than say, the manufacturing industry. Similarly, the performance requirements generally expected from telecom workloads (that can have hundreds of millions of users) is more than for most other industries. We make the reasonable assumption that the complexity quotient 
that we observed for the target managed environment also applies 
to the hybrid cloud managing environment. It is 
logical that the larger the value of the complexity quotient, the smaller
should be the value of $\delta_w$, which is consistent with our
observations. To take an example, Table~{\ref{tab:data2verifydeltaw}}
reveals that retail systems manifest a lesser complexity quotient 
than financial systems; they also empirically demonstrate a 
higher $\delta_w$ than financial systems. What this means is that, 
financial workloads yield a larger value for $H(w)$ compared to 
retail workloads for a given number and spread of workloads across 
the four Quadrants in Figure~{\ref{fig:clic}}.
\vspace{-10pt}
\subsection{Threats to Validity}
Like any other empirical study, it is important to highlight various threats to the validity while placing confidence in the results. In this section, we comment on the threats to validity of
our results under 3 different axes: \textit{construct validity}, \textit{internal validity}, and \textit{external validity}~\cite{Cook2002}.

Construct Validity implies that the dataset used for the experiment is correct. Our data is in essence, meta-data compiled while working on hundreds of client deals. Also, the data used to construct our model
 ({Table~\ref{tab:data4model}} and {Table~\ref{tab:data4deltaw}}) does not overlap with the data leveraged to verify the model ({Table~\ref{tab:data2verifymodel}}).
 Thus, while potential threats under this head are minimal, residual risks include the possibility of wrongly determining 
 $K$ or $x$ for the service provider in question, and errors while empirically calculating $\delta_w$. On the latter, our model, for the sake of simplicity, ignores the possibility of a slight variation of $\delta_w$ across the four quadrants for a given industrial grouping, and this can ripple into a slight error in predicting $\widehat E(w)$ for some corner cases. It is also possible that the best classification of a given set of workloads is not the industry that it belong to, rather its phase in the DevOps life cycle, for example, development-test. For such cases, the corresponding $\delta_w$ will need to be experimentally deciphered by the practitioner as described in Figure~{\ref{fig:delta-graph}}.

Internal validity highlights the impacts of the experimental process, conditions, past history, or relationships among inputs on the observed outcome. Our sample space spans across 60 and 90 independent cloud deployment projects as described in {Table~\ref{tab:data4model}} and {Table~\ref{tab:data4deltaw}}, respectively. Also, these projects have cross-geography, cross-industry and cross-provider spread. The tools, methodology and the data collection mechanism used for executing these deployment projects are uniform and free from any project specific biases. 

We have attempted to establish \textit{external validity} by applying our findings to twelve randomly chosen case studies across 4 industries and calculating the variance of what we observed with what our model predicted. The spread of industry domains in our chosen sample set has been done to establish the generality of the approach. While we are confident of the general applicability of our model, because of the sheer diversity of real life use cases, there could be occasional divergence from what our model predicts. As we continue to test our model on new and more complex cloud deployments, we may discover the need to render it more granular to maintain accuracy in the face of wide adoption. We will continue to refine our model with future work, whose contours we lay out in Section~{\ref{sec:concl}}.

\vspace{-10pt}
\section{Related Work}
\label{sec:relwork}
While cloud computing has been in existence for nearly a decade and
has gained attention of both academia and practitioners, the notion of
hybrid cloud which is an integration of private and public clouds, is
a recent phenomenon. As observed in~\cite{Sotomayor2009}, early
versions of hybrid clouds aimed to supplement local and private
infrastructure with compute
capabilities borrowed from the public IaaS, but this approach does
not provide a seamless integration across compute environments. 
With the growing popularity of hybrid clouds, it is
now all the more important that the public-private IaaS integration is
seamless and transparent to the user. In order to deploy a large 
enterprise software intensive system on a hybrid cloud, the software
architect must analyze and partition various workloads of the system
onto the hybrid infrastructure leveraging a methodology such as the
one we propose.

An approach by Zhang et al.~\cite{Zhang2009} discusses a factoring
algorithm that applies for Internet-based applications with dynamic
workloads. The mechanism detects and splits workloads to 'base' and
'trespassing' zones when encountering load spikes. The latter zone is
assumed to contain requests for a small number of unique data items
that are best served on elastic public clouds 
Though this work is not aimed towards a hybrid deployment, the
approach attempts to partition workloads.

Smit et al.~\cite{Smit2012} proposes a code partitioning method for
client-facing web applications that first annotates code based on
characteristics such as sensitivity to mobility. The partitioning
algorithm takes these cues and apportions it onto public or private
clouds.

Literature such as~\cite{Rosas2012} propose improvements to decrease
load imbalance in data-intensive applications running on
high-performance computing systems; their focus is on tuning the
way data is partitioned into chunks depending on execution times and
the number of nodes in the system.

The approach described in~\cite{Javadi2012} proposes a resource
provisioning policy based on the workload model so as to minimize
failure. The work proposes a scheduling infrastructure that can
suitably allocate a workload to an infrastructure.

Kaviani et al.~\cite{NimaKaviani2014} propose a method to partition
multi-tier web applications for hybrid clouds. Dependencies across the
tiers are modeled, which yields suggestions on the placement of
application and data tiers across public cloud or on-premise private
clouds.

While literature such as~\cite{Zhang2009,Smit2012,NimaKaviani2014}
have explored workload partitioning for hybrid clouds, they are
attuned to certain categories of software systems, primarily
Internet-facing web applications. We present a more generic approach
to architectural partitioning that is heuristically applicable across
wide categories of software systems across industries. We also go a
step further and propose a method to translate the architectural
partitioning into deployment complexity and associated effort
estimates. Also, our semantic of what constitutes hybrid IT deployment
is broader than prior art, which only chooses between two alternatives:
on-premise private clouds and off-premise public clouds.
\vspace{-5pt}
\section{Conclusion and Future Work}
\label{sec:concl}
In this paper, we presented a heuristic approach to help solve the
problem of rapidly determining the best-fit hybrid deployment
architecture for a given complex software system. We introduced what
we call the Cloud Line of Isolation and Control (CLIC) that serves as
a high-level deployment-centric partitioning indicator to segment
workloads based on their characteristics and requirements. We went on
to propose a mathematical model to represent hybrid deployments, and 
a metric to measure the degree of hybrid complexity of categories 
of workloads. We then used this model to develop a method to predict 
the effort  estimate to deploy and sustain hybrid cloud environments.We also 
drew industry-specific conclusions from data collected from various client
deals across industries.

We next evaluated our model using data from twelve case studies across four 
industries to which we applied the hybrid cloud complexity model that we constructed. 
We measured the variance of predicted and observed results and verified that our model 
is a reasonably accurate tool to forecast the complexity and effort estimate to implement 
and run hybrid cloud environments.

Our approach however, has limitations. The generality of applicable
workloads and the perfection of partitioning are complementary
variables, so there is a fundamental limit to the precision with which
both can be achieved simultaneously. In future, we plan to improve our
approach by enhancing the coverage of many deployment possibilities in
a hybrid hosted environment. For example, the four quadrants of
Figure~\ref{fig:clic} can be refined to nine (three on each axis
rather than two); and there can be multiple sub-CLIC lines that
separate more of the resulting zones. Additionally, $\delta_w$, now 
generalized across the deployment quadrants, could be rendered quadrant-specific. 
And what constitutes a workload can be laid out in a more granular fashion in terms of underpinning servers, middleware tiers and application components.
The equations that we have derived to measure hybrid complexity can also correspondingly
mature. Furthermore, our complexity and effort estimate models are designed for the managing
environment (the cloud stack) and not the managed environment. The
complexity of virtualizing and sustaining the managed environment depends on the nature of
workloads and the composition of the compute, storage and network domains in the data center; 
it will also need to factor in aspects such as workload migration
which we have not considered in this paper.

The emerging area of Cloud Service Brokerage~\cite{Plummer2012} is
expected to mitigate many problems attendant with consuming
hybrid IT. 
Cloud brokerage can create and support constructs that abstract
underpinning clouds, thereby delivering an integrated hybrid IT experience. 
We also intend to focus on this area for future
research on hybrid Clouds.
%
%
%\vspace{-5pt}
\small{
\bibliographystyle{abbrv}
%\vspace{-5pt}
\bibliography{cloud.krish}
}
\end{document}